\documentclass[12pt]{article}
\usepackage{latexsym}
\usepackage{amsmath}
\usepackage{amsfonts}
\usepackage{amssymb}
\usepackage{amscd}
\usepackage{fancybox}
\usepackage{cite}
\usepackage{amsmath,amsfonts,amsbsy}
\usepackage{pstricks,pst-node}
\usepackage[small,bf,hang]{caption2}

\usepackage[vcentermath]{youngtab}

\usepackage{float}

\psset{unit=1.3cm,linewidth=.5pt,radius=.2}  

\newcommand{\fund}{{\tiny \yng(1)}}
\newcommand{\twoform}{{\tiny \yng(1,1)}}
\newcommand{\threeform}{{\tiny \yng(1,1,1)}}
\newcommand{\fourform}{{\tiny \yng(1,1,1,1)}}

\newcommand{\symm}{{\tiny \yng(2)}}

\usepackage[vcentermath]{youngtab}          
\usepackage{enumitem}                       
\usepackage{multirow}                     
\usepackage{float}                          
\usepackage{lscape}                         
\usepackage{bm}

\def\hybrid{\topmargin -20pt    \oddsidemargin 0pt
        \headheight 0pt \headsep 0pt
        \textwidth 6.25in       
        \textheight 9.5in       
        \marginparwidth .875in
        \parskip 5pt plus 1pt   \jot = 1.5ex}

\hybrid

\newcommand{\beq}{\begin{equation}}
\newcommand{\eeq}{\end{equation}}
\newcommand{\bi}{\begin{itemize}}
\newcommand{\ei}{\end{itemize}}
\newcommand{\bea}{\begin{eqnarray}}
\newcommand{\eea}{\end{eqnarray}}
\newcommand{\ba}{\begin{array}}
\newcommand{\ea}{\end{array}}
\newcommand{\bt}{\begin{tabular}}
\newcommand{\et}{\end{tabular}}
\newcommand{\bc}{\begin{center}}
\newcommand{\ec}{\end{center}}

\newcommand{\ft}[2]{{\textstyle {\frac{#1}{#2}} }}

\begin{document}

\begin{titlepage}
\begin{center}

\hfill UG-08-05 \\

\vskip 1.5cm

{\Large \bf Multiple M2-branes and the Embedding Tensor
\\[0.2cm]}

\vskip 1.5cm

{\bf Eric A.~Bergshoeff, Mees de Roo and Olaf Hohm} \\

\vskip 25pt

{\em Centre for Theoretical Physics, University of Groningen, \\
Nijenborgh 4, 9747 AG Groningen, The Netherlands \vskip 5pt }

{email: {\tt E.A.Bergshoeff@rug.nl, M.de.Roo@rug.nl, O.Hohm@rug.nl}} \\

\vskip 0.8cm

\end{center}

\vskip 1cm

\begin{center} {\bf ABSTRACT}\\[3ex]

\begin{minipage}{13cm}
\small We show that the Bagger-Lambert theory of multiple M2-branes
fits into the general construction of maximally supersymmetric gauge
theories using the embedding tensor technique. We apply the
embedding tensor technique in order to systematically obtain the
consistent gaugings of ${\cal N}=8$ superconformal theories in $2+1$
dimensions. This leads to the Bagger-Lambert theory, with the
embedding tensor playing the role of the four-index antisymmetric
tensor defining a ``3-algebra''. We present an alternative
formulation of the theory in which the embedding tensor is replaced
by a set of unrestricted scalar fields. By taking these scalar
fields to be parity-odd, the Chern-Simons term  can be made
parity-invariant.

\end{minipage}

\end{center}

\noindent

\vfill

April 2008

\end{titlepage}

\section{Introduction}\setcounter{equation}{0}

Recently, a three-dimensional world-volume theory describing a set
of multiple M2-branes was proposed \cite{Bagger:2006sk,
Bagger:2007jr, Bagger:2007vi} (see also
\cite{Gustavsson:2007vu,Gustavsson:2008dy}).
The theory is based on the existence of a ``3-algebra'' that
generalizes the Lie algebras of ordinary gauge theories to a
structure involving an antisymmetric triple bracket
 \bea
  [T^{a},T^{b},T^{c}] \ = \ f^{abc}{}_d T^{d}\;,
 \eea
where $T^{a},\ a=1,\ldots,N$, denote the generators of the algebra.
Assuming the existence of a symmetric tensor $h_{ab}$ (which we will
later take to be the Kronecker delta $\delta_{ab}$) to raise and
lower indices, the generalized structure constants $f^{abcd}$ need
to be totally antisymmetric,
\begin{equation} \label{linear1}
f^{abcd} = f^{[abcd]}\;.
\end{equation}
Apart from this linear identity, there is a quadratic identity (the
so-called fundamental identity),\footnote{Such structures also occur
in the study of maximally supersymmetric solutions of supergravity
theories \cite{FigueroaO'Farrill:2002xg}.}
\begin{equation}\label{quadratic1}
f^{abe}{}_g f^{cdg}{}_f - f^{cde}{}_g f^{abg}{}_f - f^{abc}{}_g
f^{dge}{}_f + f^{abd}{}_g f^{cge}{}_f =0 \,,
\end{equation}
which is the analogue of the Jacobi identity for Lie algebras.
Sofar, only one explicit solution of the constraints \eqref{linear1}
and \eqref{quadratic1} is known, namely
$f^{abcd}=\varepsilon^{abcd}$ for $N=4$. In this case it has been
shown that the theory can be reinterpreted as an ordinary gauge
theory based on the gauge group $SO(4)=SU(2)\times SU(2)$
\cite{VanRaamsdonk:2008ft}.

The world-volume theory describing multiple M2-branes
contains a set of embedding scalars $X_a^I,\
 I=1,\ldots,8$ and a set of fermions $\Psi_a$.
Furthermore, there is a set of world-volume gauge fields $A_{\mu
ab}=-A_{\mu ba}$. A crucial feature of the theory is that these
gauge fields do not describe independent degrees of freedom. They
occur via a Chern-Simons term such that their field equations lead
to a duality relation with the embedding scalars. This Chern-Simons
term was introduced in an earlier attempt to construct a
supersymmetric world-volume theory with 16 supercharges
\cite{Schwarz:2004yj}. A nice feature of the Bagger-Lambert theory
is that it reproduces the so-called Basu-Harvey equation
\cite{Basu:2004ed} which was the original motivation for the
proposal of \cite{Bagger:2006sk, Bagger:2007jr, Bagger:2007vi}.

The above two features: (1) a tensor that satisfies a linear and a
quadratic constraint and (2) gauge fields that occur via a
Chern-Simons term, are very reminiscent of the so-called embedding
tensor technique for constructing matter-coupled gauged
supergravities. This method was originally proposed to construct
maximal gauged supergravities in three dimensions
\cite{Nicolai:2000sc,Nicolai:2001sv} and later applied to other
cases in three dimensions \cite{Nicolai:2001ac,deWit:2003ja,Schon:2006kz}. The
embedding tensor $\Theta$ plays the role of the tensor $f$ above and
is used to specify which gauge fields are needed to gauge which
subgroup of the duality group.

The case of matter-coupled half-maximal supergravity in three
dimensions, with duality group $SO(8,N)$,  was studied in
\cite{Nicolai:2001ac,deWit:2003ja,Schon:2006kz}. The relevant
embedding tensor is a 4-index tensor $\Theta_{ab,cd}$ satisfying
certain linear and quadratic constraints. A particular solution to
these constraints is given by $\Theta_{ab,cd}=f_{abcd}$ satisfying
(\ref{linear1}) and (\ref{quadratic1}). However, in supergravity
there are more possibilities than this totally antisymmetric
combination. Specifically, $\Theta$ can have a singlet
(corresponding to a gauging of the full duality group) and a
symmetric traceless part.

Sofar, the embedding tensor technique has been mainly applied to
construct gauged supergravity theories but it can be used to
construct supersymmetric gauge theories as well. For instance, it
has been used to construct ${\cal N}=2,D=4$ supersymmetric gauge
theories with electric and magnetic charges  \cite{deVroome:2007zd}.
In this note we wish to apply the embedding tensor technique to the
case of ${\cal N}=8$ supersymmetric gauge theories in three
dimensions and find out whether generalizations of the
Bagger-Lambert model are possible or not. This investigation is also
a nice illustration of how the embedding tensor technique works
in general.

\section{Gauging  ${\cal N}=8$ superconformal theories}
Our starting point is the free superconformal ${\cal N}=8$ theory in
$D=3$ with $N$ matter multiplets, i.e.~containing $8N$ scalars
$X^{aI}$ and $8N$ Majorana spinors $\Psi^{aA}$. Here and in the
following $I,J=1,\ldots ,8$, $A,B=1,\ldots,8$ and
$\dot{A},\dot{B}=1,\ldots,8$ denote vector, spinor and conjugate
spinor indices of the $SO(8)$ R-symmetry group. The theory is
described by the Lagrangian
 \bea
  {\cal L} \ = \ -\ft12 \partial^{\mu}X^{aI}\partial_{\mu}X_a^{I}
  +\tfrac{i}{2}\bar{\Psi}^{aA}\Gamma^{\mu}\partial_{\mu}\Psi_a^{A}\;.
 \eea
It is invariant under the supersymmetry transformations\footnote{We
suppress the $SO(8)$ spinor indices whenever they are not explicitly
required.} (with $\Gamma_{012}\epsilon = \epsilon$)
 \bea
  \delta_{\epsilon}X^{I}_a \ = \ i\bar{\epsilon}\,\Gamma^{I}\Psi_a\;,
  \qquad
  \delta_{\epsilon}\Psi_a \ = \
  \partial_{\mu}X^{I}_a\Gamma^{\mu}\Gamma^{I}\epsilon\;,
 \eea
and under the global symmetry group $SO(8)\times SO(N)$.

We wish to analyze the question which subgroups of this global
symmetry group can be promoted to a local symmetry. Since in this
note our ultimate motivation is the M2 brane example, we will
restrict ourselves to gauge groups that lie inside the $SO(N)$
factor. As usual, we first introduce gauge-covariant derivatives
that couple the scalars to the gauge vectors. As in
\cite{Nicolai:2000sc,Nicolai:2001ac} these gauge fields come in the
adjoint representation of the global symmetry group $G_{\rm global}$
and enter only via a Chern-Simons term. The covariant derivatives
read
 \bea\label{cov}
   D_{\mu} \ = \
   \partial_{\mu}-g\,\Theta_{\alpha\beta}A_{\mu}{}^{\alpha}t^{\beta}\;,
 \eea
where $g$ is the gauge coupling constant and the indices
$\alpha,\beta=1,\ldots,{\rm dim}\hspace{0.1em}G_{\rm global}$ label
the adjoint of the rigid symmetry group, spanned by the generators
$t^{\alpha}$ with structure constants $f^{\alpha\beta}{}_{\gamma}$.
The symmetric embedding tensor
$\Theta_{\alpha\beta}=\Theta_{\beta\alpha}$ encodes the embedding of
the gauge group $G_0$ into $G_{\rm global}$ in that $G_0$ is spanned
by generators
 \bea\label{gaugealg}
  X_{\alpha} \ = \ \Theta_{\alpha\beta}t^{\beta}\;.
 \eea
In other words, $\Theta$ acts as a projector which singles out those
generators that participate in the gauging. Gauge invariance of the
theory to be constructed requires invariance of
$\Theta_{\alpha\beta}$ under the adjoint action of the gauge group
generators $X_{\alpha}$. This implies the quadratic constraint
 \bea\label{quadconstr}
  {\cal Q}_{\alpha,\beta\gamma} \ \equiv \
  \Theta_{\alpha\epsilon}\Theta_{\delta(\beta}f^{\delta\epsilon}{}_{\gamma)}
  \ = \ 0\;,
 \eea
which also ensures closure of the gauge algebra spanned by
(\ref{gaugealg}).

In the case at hand, the indices split according to the adjoint of
$SO(N)$, i.e.~$\alpha =[ab]$. Consequently, the embedding tensor
reads $\Theta_{ab,cd}$ and has the symmetries
\begin{equation} \label{linear2}
\Theta_{ab,cd} = -\Theta_{ba,cd} = - \Theta_{ab,dc} =
\Theta_{cd,ab}\;.
\end{equation}
Using the explicit form of the structure constants
 \bea\label{structure}
  f^{ab,cd,}{}_{ef} \ = \
  -2\delta^{[a}{}_{[e}\delta^{b][c}\delta^{d]}{}_{f]}\;,
 \eea
the quadratic constraint reads
 \bea\label{quadraticAd}
  {\cal Q}_{ab,cd,ef} \ \equiv \ \Theta_{ab,e}{}^g\Theta_{cd,gf} -
  \Theta_{cd,e}{}^g\Theta_{ab,gf} - \Theta_{ab,c}{}^g\Theta_{dg,ef}
  +\Theta_{ab,d}{}^g\Theta_{cg,ef} \ = \ 0\;.
 \eea
Moreover, the generators in (\ref{cov}) act in the fundamental
representation, $(t^{ab})^{c}{}_d=\delta^{[a}{}_d\delta^{b]c}$, i.e.
the explicit form of the covariant derivative is given by
\begin{equation}
D_\mu X_d^I = \partial_\mu X_d^I
-gA_\mu{}^{ab}\,\Theta_{ab,cd}\,X^{cI}\;,
\end{equation}
and similarly for the spinors.

Let us now turn to the gauged action. Our starting point is the
following Ansatz
\begin{eqnarray} \label{lagr}
{\cal L}_g &=& -\tfrac{1}{2} D^\mu X^{aI}D_\mu X_a^I
+\tfrac{i}{2}\bar\Psi^a \Gamma^\mu D_\mu\Psi_a
+\tfrac{i}{4}g\bar\Psi^{aA}A_{3\hspace{0.1em}aA,bB}(X)\Psi^{bB} \\
\nonumber &&+
\tfrac{1}{2}g\,\varepsilon^{\mu\nu\lambda}A_{\mu}{}^{\alpha}\Theta_{\alpha\beta}
\left(\partial_{\nu}A_{\lambda}{}^{\beta}-\tfrac{1}{3}g\Theta_{\gamma\delta}f^{\beta\delta}{}_{\epsilon}
A_{\nu}{}^{\gamma}A_{\lambda}{}^{\epsilon} \right)-g^2 V(X)\,.
\end{eqnarray}
As in supergravity, we added Yukawa-like couplings parameterized by
a scalar-dependent function $A_3(X)$ as well as a scalar potential
$V(X)$ and a Chern-Simons term. By virtue of the quadratic
constraint (\ref{quadconstr}) this action is gauge invariant under
 \begin{eqnarray}
  \delta X_a^{I} &=&
  -g\Lambda^{\alpha}\Theta_{\alpha\beta}(t^{\beta})^{b}{}_{a}X_{b}^{I}
  \ = \ g\Lambda^{cd}\Theta_{cd,ba}X^{bI}\;, \\ \nonumber
  \delta \Psi_a &=& g\Lambda^{cd}\Theta_{cd,ba}\Psi^b\;, \\
  \nonumber
  \delta A_{\mu}{}^{\alpha} &=& D_{\mu}\Lambda^{\alpha} \ = \ \partial_{\mu}\Lambda^{\alpha}
  -g\Theta_{\beta\gamma}f^{\alpha\beta}{}_{\delta}A_{\mu}{}^{\gamma}\Lambda^{\delta}\;.
 \end{eqnarray}
For the gauge vectors $A_{\mu}{}^{ab}$ with explicit $SO(N)$ indices
the gauge variation can be rewritten by use of the structure
constants (\ref{structure}) as
 \bea
  \delta A_{\mu\hspace{0.1em} ab} \ = \ \partial_{\mu}\Lambda_{ab}
  +g\Theta_{ae,cd}A_{\mu}{}^{cd}\Lambda^{e}{}_{b}
  -g\Theta_{be,cd}A_{\mu}{}^{cd}\Lambda^{e}{}_{a}\;.
 \eea

Next, we are going to analyze the question for which choices of the
embedding tensor the action corresponding to (\ref{lagr}) can be
made supersymmetric. We use the following Ansatz for the
supersymmetry variation of the fermions
 \bea
   \delta_{\epsilon}\Psi_{aA} \ = \
   D_{\mu}X_a^{I}\,\Gamma^{\mu}\Gamma^{I}_{A\dot{B}}\,\epsilon^{\dot{B}}
   +gA_{2\hspace{0.1em}aA\dot{B}}(X)\,\epsilon^{\dot{B}}\;,
 \eea
where we introduced a gauge-covariant derivative and added a
scalar-dependent fermion shift function $A_2(X)$. Due to the
non-commutativity of covariant derivatives, the supersymmetry
variation of the kinetic terms in (\ref{lagr}) no longer vanishes,
but instead gives rise to a term proportional to the field strength,
 \bea\label{kinvar}
  \delta_{\epsilon}{\cal L}_{\rm kin} \ = \
  \tfrac{i}{2}g\Theta_{ab,cd}\bar{\Psi}^{a}\Gamma^{\mu\nu}\Gamma^{I}\epsilon
  F_{\mu\nu}{}^{cd}X^{bI}\;.
 \eea
These can be compensated by assigning a non-trivial supersymmetry
variation to the gauge vectors,
 \bea
  \delta_{\epsilon}A_{\mu}{}^{ab} \ = \
  i\bar{\epsilon}\Gamma_{\mu}\Gamma^{I}X^{I[a}\Psi^{b]}\;,
 \eea
such that the variation of the Chern-Simons term precisely cancels
(\ref{kinvar}). However, these variations of the gauge field give
rise to additional variations from its presence inside the covariant
derivatives, and the problem is to determine $A_2(X)$ and $A_3(X)$
such that these contributions can be canceled.

In the embedding tensor formalism this problem of finding a
consistent supersymmetric deformation translates into the problem of
finding the right linear constraints on the embedding tensor. A
priori, $\Theta_{ab,cd}$ with the symmetries (\ref{linear2}) takes
values in the symmetric tensor product
 \bea
  \left(\hspace{0.2em}{\tiny \yng(1,1)}\otimes {\tiny \yng(1,1)}\hspace{0.2em}\right)_{\rm sym}
  \ = \  {\bf 1} \oplus {\tiny \yng(1,1,1,1)} \oplus
  {\tiny \yng(2,2)}\oplus {\tiny \yng(2)}\;.
 \eea
This corresponds to the general parametrization
 \begin{equation}\label{linear3}
  \Theta_{ab,cd} \ = \ f \delta_{c[a}\delta_{b]d}
  +f_{abcd} + f^{\tiny (2,2)}_{ac,bd}+ f_{[c[a}\delta_{b]d]}\;,
 \end{equation}
where $f_{abcd}=f_{[abcd]}$ denotes the totally antisymmetric part
and $f^{(2,2)}_{ac,bd}$ has the window symmetries of ${\tiny
\yng(2,2)}$, in particular $f^{(2,2)}_{[ac,b]d}=0$. Supersymmetry
then implies that only some of these irreducible representations are
consistent, or in other words, beyond the quadratic constraint
(\ref{quadraticAd}) it requires a $G_{\rm global}$--covariant linear
constraint.

To determine these constraints, we first focus on the variations
linear in the gauge coupling $g$ (and thus $\Theta$) and linear in
the fermions $\Psi$. The variation $\delta A_{\mu}{}^{ab}$ inside
the covariant derivative on the scalars gives rise to terms of the
form $X^2 DX\Psi$. In order to cancel these it follows that $A_3(X)$
has to be quadratic in $X$ and, consequently, $A_2(X)$ has to be
cubic in $X$. The most general Ansatz in terms of the embedding
tensor reads
 \begin{eqnarray}\label{A2ansatz}
  A_{3\hspace{0.1em}aA,bB} &=&
  b_1\Theta_{ab,cd}\Gamma^{IJ}_{AB}X_c^{I}X_d^{J}
  +b_2\Theta_{ac,bd}\delta_{AB}X^{I}_cX^{I}_{d}
  +b_3\Theta_{ac,bd}\delta^{cd}\delta_{AB}X_{e}^{I}X_{e}^{I}\;, \\
  \nonumber
  A_{2\hspace{0.1em}aA\dot{B}} &=&
  \Theta_{ab,cd}\left(c_1\Gamma^{IJK}_{A\dot{B}}
  X^{Ib}X^{Jc}X^{Kd}
  +c_2\Gamma^{I}_{A\dot{B}}X^{Id}X^{Jb}X^{Jc}\right)\;.
 \end{eqnarray}
The variation of the Lagrangian gives rise to a term proportional to
$b_1\Gamma^{IJ}\Gamma^{K}$, containing the antisymmetric part
$\Gamma^{IJK}$. These have to be canceled by choosing the
coefficient $c_1$ in $A_2$ and thus in the supersymmetry variation
of the fermion in the right way. However, from (\ref{A2ansatz}) one
infers that this term in $A_2$ can only be non-zero if
$\Theta_{[ab,c]d}$ is non-zero. This in turn implies that only the
totally antisymmetric $f_{abcd}$ in (\ref{linear3}) can give rise to
a consistent gauging. Specifically one finds
 \bea
  b_1 =-1\;, \quad b_2=b_3=0\;, \qquad c_1 =\ft16\;, \quad
  c_2=0\;
 \eea
and the following expression for the scalar potential $V(X)$:
 \begin{equation}
  V(X) = \tfrac{1}{12} \Theta_{ab,cg}\,\Theta_{de,f}{}^g\, X^{aI}
  X^{bJ}
  X^{cK} X^{dI} X^{eJ} X^{fK}\;.
 \end{equation}
In total, the linear constraint imposed by supersymmetry reads
 \bea\label{linconst}
  \left(\mathbb{P}_{{\bf 1} \hspace{0.2em}\oplus\hspace{0.2em}
  {\tiny \yng(2,2)}\hspace{0.2em}\oplus\hspace{0.2em} {\tiny \yng(2)}}\right)\Theta_{ab,cd} \ = \
  0 \;,
 \eea
 where $\mathbb{P}$ projects out those representations that are not totally antisymmetric.
A particular solution, for $N=4$,  is given by
$\Theta_{ab,cd}=\varepsilon_{abcd}$, which is an invariant tensor of
$SO(4)$ and solves the quadratic constraints (\ref{quadraticAd}).
This leads to the $SO(4)$ gauge theory example of
\cite{Bagger:2006sk, Bagger:2007jr, Bagger:2007vi}.

The general solution $\Theta_{ab,cd}=f_{abcd}$ of (\ref{linconst})
gives  back the Bagger-Lambert theory. In fact, for a totally
antisymmetric embedding tensor the quadratic constraint
(\ref{quadraticAd}) precisely reduces to the fundamental identity
(\ref{quadratic1}). Moreover, all the couplings match. In particular
the Chern-Simons term of \cite{Bagger:2007jr} based on the 3-algebra
structure constants precisely coincides with the Chern-Simons action
in \eqref{lagr}, as can be checked by insertion of
(\ref{structure}).

In order to illustrate the use of the embedding tensor, let us
briefly comment on different choices of gauge groups and their
embedding. First of all, for any $N\geq 4$ the linear constraint
(\ref{linconst}) allows for a consistent gauging of the subgroup
$SO(4)\subset SO(N)$. Splitting the indices as $a=(i,5,\ldots,N)$,
this corresponds to a choice of embedding tensor, in which the only
non-vanishing components are
 \bea
  \Theta_{ij,kl} \ = \
  \varepsilon_{ijkl}\;.
 \eea
This is an invariant tensor of the subgroup to be gauged, and it
solves the quadratic constraint. Depending on the value of $N$,
larger gauge groups may be possible. For instance, for $N=8$ the
canonically embedded $SO(4)\times SO(4)$ subgroup can be gauged,
corresponding to an embedding tensor of the form
 \bea
  \Theta_{ij,kl} \ = \
  \kappa_1\,\varepsilon_{ijkl}\;, \qquad
  \Theta_{\bar{i}\bar{j},\bar{k}\bar{l}} \ = \
  \kappa_2\,\varepsilon_{\bar{i}\bar{j}\bar{k}\bar{l}}\;.
 \eea
Here we have split the indices according to
$a=(i,\bar{i})=(1,\ldots,4,\bar{1},\ldots,\bar{4})$ and introduced
two arbitrary coupling constants $\kappa_1$, $\kappa_2$.

One may wonder whether more interesting gauge groups are possible,
beyond the various copies of $SO(4)$. An attractive candidate is
$G_2$ in case of $N\geq 7$. In fact, it can be defined as the
subgroup of $SO(7)$ that leaves a certain antisymmetric 4-tensor
$C_{abcd}$ invariant, and so one may take
$\Theta_{ab,cd}=f_{abcd}=C_{abcd}$. (For a concise account of $G_2$
see, for instance, appendix A of \cite{Bilal:2001an}.) This ansatz
has been pursued in \cite{Bandres:2008vf} (see also
\cite{Ho:2008bn}), with negative results. The fundamental identity
(\ref{quadratic1}) is not satisfied and so $G_2$ does not give rise
to a consistent 3-algebra. It is, however, instructive to reexamine
this problem from the point of view of the embedding tensor
formalism. From this perspective, the embedding tensor should act as
a projector from $SO(7)$ onto $G_2$ according to (\ref{gaugealg}).
It is possible to find such a projector which is totally
antisymmetric, i.e., satisfying the linear constraint, and which
gives rise to the closed $G_2$ algebra. However, this tensor is not
$G_2$ invariant and therefore the quadratic constraints
(\ref{quadraticAd}) are still not satisfied. Instead one may start
from the 4-tensor of $G_2$ which is known to be invariant. It is
given by
 \bea\label{4tensor}
  C_{abcd} \ = \ \tfrac{1}{3!}\varepsilon^{abcdefg}\hat{C}_{efg}\;,
 \eea
where $\hat{C}_{abc}$ is defined by
 \bea
  \hat{C}_{ijk} = \varepsilon_{ijk}\;, \qquad
  \hat{C}_{i\bar{j}\bar{k}} = \hat{C}_{\bar{i}j\bar{k}}=
  \hat{C}_{\bar{i}\bar{j}k}=-\varepsilon_{ijk}\;, \qquad
  \hat{C}_{7 i\bar{j}}=\delta_{ij}\;,
 \eea
and we have split the indices according to $a=(i,\bar{i},7)$.
However, (\ref{4tensor}) is not a projector onto $G_2$. A possible
solution for the embedding tensor resp.~the projector is instead
given by
 \bea\label{Thetasol}
  \Theta_{ab,cd} \ = \ \delta_{c[a}\delta_{b]d}+\ft14 C_{abcd}\;.
 \eea
By insertion of (\ref{Thetasol}) into (\ref{quadraticAd}) one can
verify that this indeed solves the quadratic constraints. But it
does not solve the linear constraint (\ref{linconst}) due to the
presence of a singlet combination. However, as we will discuss in
the next section, in gauged supergravity these singlet components
are allowed \cite{Nicolai:2001sv,deWit:2003ja}. Therefore we
conclude that while the superconformal theories do not allow for a
gauging of $G_2$, the embedding tensor (\ref{Thetasol}) does give
rise to a consistently gauged ${\cal N}=8$ supergravity. To the best
of our knowledge this example has not appeared in the literature
before.

\section{Comparing with Gauged Supergravity}

In this section we will compare the application of the embedding
tensor technique to both half-maximal matter-coupled gauged
supergravities as well as to maximal supersymmetric gauge theories.
Both theories have an equal number of supercharges. In this section
we will not only consider $D=3$ but also $3<D \leq 10$ dimensions.
It turns out that the supergravity theories allow for more
consistent gaugings. This is due to the fact that these theories
have a weaker linear constraint and a less trivial (non-compact)
duality group to start with.

It is of interest to compare the embedding tensors corresponding to
the supergravity and gauge theory cases in more detail. At first
sight the two gaugings are unrelated since the relevant scalar
manifolds differ. For instance,  the $D=3$ supergravity case leads
to the scalar manifolds $SO(8,N)/SO(8)\times SO(N)$ whereas  in the
$D=3$ gauge theory case one deals  with the flat manifolds
$\mathbb{R}^{8N}$. Nevertheless, we will argue below that the
embedding tensor representations for the gauge theories can be
deduced from the corresponding supergravity representations. For the
supergravity case the representations of the
 embedding tensors have been calculated
 \cite{Nicolai:2001ac,deWit:2003ja,Schon:2006kz,Bergshoeff:2007vb},
 see table 1.\footnote{We have indicated in the table only gaugings, no massive deformations.
Furthermore, we have ignored the chiral case in $D=6$ dimensions.
For more details, see table 3 in \cite{Bergshoeff:2007vb}.}
\begin{table}[t]
\begin{center}
    \begin{tabular}{|c||c||c|}
    \hline
   $D$&supergravity&gauge theory\\
    \hline
    \hline
    &&\\ [-10pt]
    10 -- 6  & $\fund \;\;\;\;\threeform$   &$\threeform$\\  [10pt]
    \hline && \\ [-10pt]
     5           & $\fund \;\;\;\; \twoform \;\;\;\; \threeform$     &$\threeform$\\ [10pt]
    \hline && \\ [-10pt]
    4           & $\left(\bf 2,\fund \right)\;\;\; \; \left(\bf 2, \threeform \, \right)$ &
    $\left(\bf 2, \threeform \, \right)$ \\ [10pt]
    \hline && \\ [-10pt]
    3           & $1 \;\;\; \symm \;\;\;\; \fourform$    &   $\fourform$  \\ [15pt]
    \hline
    \end{tabular}
    \caption{This table gives, for $3\le D\le 10$, the duality group representations of the embedding tensors for
matter-coupled half-maximal gauged supergravity (second column) and
maximal supersymmetric gauge theory (third column). In $D=4$ the
${\bf 2}$ refers to the fundamental representation of the
electro-magnetic $SL(2,\mathbb{R})$ duality}\label{halfmax}
\end{center}
\end{table}
We see that for supergravity in $6\le D \le 10$ dimensions there is
a fundamental and three-form representation. In these dimensions the
fundamental representation corresponds to the gauging of a $SO(1,1)$
diagonal subgroup of the $SO(1,1)\times SO(10-D,10-D+N)$ duality
group\,\footnote{Actually, the quadratic constraints forbid this
gauging in $D=10$ because $SO(N)$ has no $SO(1,1)$ subgroup.}. The
three-form representation represents the anti-symmetric structure
constants of the subgroup $G_0 \subset SO(10-D,10-D+N)$ that is
gauged. The reason that the fundamental representation is absent in
the gauge theory case is that the corresponding $SO(1,1)$ symmetry
that is gauged involves a shift of the dilaton supergravity field,
which is absent in the gauge theory.

To be able to do more general gaugings one needs more space-time
vectors than only the fundamental representation, which is present
in all these dimensions. For example, in $D=5$ an additional vector
is provided by the dual of the NS-NS two-form, giving rise to an
extra two-form representation of possible gaugings. Since this extra
possibility is due to the dualization of a supergravity field, this
extra two-form representation is absent in the gauge theory. In
$D=4$ the extra vectors are the Hodge duals of the original ones,
leading to an $SL(2,\mathbb{R})$ doublet of possible gaugings. This
possibility arises both in the supergravity and in the gauge theory
case. In the gauge theory case, these correspond to the electric and
magnetic gaugings discussed, for ${\cal N}=2$ supersymmetry, in
\cite{deVroome:2007zd}. Finally, in $D=3$ further gaugings become
available due to the fact that scalars become dual to vectors. Note
that the singlet and symmetric traceless representation, present in
supergravity,
 have their
origin in the four-dimensional $\left(\bf 2,\fund \right)$
representation which is absent in the gauge theory. That is why the
$1$ and $\symm$ representations are absent in the $D=3$ gauge theory.

Summarizing, by comparing with half-maximal supergravity we obtain a
natural prediction for the embedding tensors of maximal
supersymmetric gauge theories as presented in table 1. This includes
the four-index anti-symmetric representation of \cite{Bagger:2006sk,
Bagger:2007jr, Bagger:2007vi}.

\section{An alternative Formulation}
Sofar, we have introduced a Lagrangian containing a constant
embedding tensor $\Theta_{ab,cd}$ satisfying the linear constraint
(\ref{linconst}) and the quadratic constraints \eqref{quadraticAd}.
Even though in this formalism the gauging takes a completely
covariant form with respect to $G_{\rm global}$, this group is no
longer an invariance of the Lagrangian. In fact, the
$\Theta_{ab,cd}$ are not dynamical objects and therefore cannot
transform under the symmetry. Instead, following
\cite{deWit:2008ta,Bergshoeff:2008qd}, we can promote the embedding
tensor to a set of unconstrained scalar fields $\Theta_{ab,cd}(x)$
with the same symmetry properties  by introducing two kinds of
Lagrange multipliers. The first set consists of 2-form potentials
$B_{\mu\nu}{}^{ab,cd}$ with the same symmetry properties as
$\Theta$. These can be viewed as the duals of $\Theta$ (see below)
and their field equations will impose the constancy of $\Theta$. The
second set consists of 3-form potentials
$C_{\mu\nu\rho}{}^{ab,cd,ef}$ which will impose the quadratic
constraints by their equations of motion. They have the same
symmetry properties as the quadratic constraint tensor ${\cal
Q}_{ab,cd,ef}$ defined in \eqref{quadraticAd}. The total Lagrangian
is then given by
\begin{equation}
{\cal L}_{\text{total}} = {\cal L}_g -\ft12 g\,
\varepsilon^{\mu\nu\rho}\,
\partial_\mu\Theta_{ab,cd}\, B_{\nu\rho}{}^{ab,cd} -\ft13 g^2\,
\varepsilon^{\mu\nu\rho}\, {\cal Q}_{ab,cd,ef}\,
C_{\mu\nu\rho}{}^{ab,cd,ef}\;,
\end{equation}
where we replaced in ${\cal L}_g$, see eq.~\eqref{lagr}, everywhere
$\Theta$ by the space-time dependent $\Theta(x)$.

The Lagrangian ${\cal L}_g$ is not gauge-invariant, neither is its
action supersymmetric\,\footnote{Of course, we should also in the
transformation rules replace $\Theta$ by its space-time dependent
form.}. However, the violation of these symmetries is proportional to
$\partial\Theta$ or to the quadratic constraint tensor ${\cal Q}$.
Such terms can always be canceled by assigning appropriate gauge
transformations and supersymmetries to the Lagrange multipliers $B$
and $C$. To illustrate this, we give the full bosonic gauge
transformations \cite{Bergshoeff:2008qd}, for which we find
\begin{eqnarray}\nonumber
\delta_{\Lambda} B_{\mu\nu}{}^{ab,cd} &=&
\partial_{[\mu}\Lambda_{\nu]}{}^{ab,cd} + D_{[\mu}\Lambda^{ab}
A_{\nu]}{}^{cd}-2\varepsilon_{\mu\nu\rho}\Lambda^{ab}X^{Ic}D^{\rho}X^{dI}\\
\nonumber&&+i\Lambda^{ab}\bar{\Psi}^c\Gamma_{\mu\nu}\Psi^d\
+\tfrac{8}{3}g\left(\Theta_{ef,g}{}^{d}\Lambda_{\mu\nu}{}^{ab,ef,cg}
-\Theta_{ef,g}{}^{c}\Lambda_{\mu\nu}{}^{ef,ab,gd}\right) \,, \cr
&&\cr \delta_{\Lambda} C_{\mu\nu\rho}{}^{ab,cd,ef} &=&
\partial_{[\mu} \Lambda_{\nu\rho]}{}^{ab,cd,ef} -
2A_{[\mu}{}^{ab}A_{\nu}{}^{cd}D_{\rho]}\Lambda^{ef}
-\varepsilon_{\mu\nu\rho}A_{\sigma}{}^{ab}\Lambda^{cd}X^{eI}D^{\sigma}X^{fI}\\
&&+3iA_{[\mu}{}^{ab}\bar{\Psi}^c\Gamma_{\nu\rho]}\Psi^d\Lambda^{ef}
+\ft18 i\varepsilon_{\mu\nu\rho}\Lambda^{ab}\bar{\Psi}^c\Gamma^{IJ}\Psi^d X^{eI} X^{fJ} \\
\nonumber &&+\tfrac{1}{24}g\Theta_{gh,i}{}^f
\varepsilon_{\mu\nu\rho}\Lambda^{ab}
X^{Ic}X^{Jd}X^{Ke}X^{Ig}X^{Jh}X^{Ki}\\ \nonumber
&&-\tfrac{1}{24}g\Theta_{gh,i}{}^{f}\varepsilon_{\mu\nu\rho}\Lambda^{ab}X^{Ig}X^{Jh}X^{Ki}X^{Ic}X^{Jd}X^{Ke}\;.
\end{eqnarray}
Here we have left implicit the Young projection of the right-hand
sides according to the symmetries of the left-hand sides. We will
not give the supersymmetry transformations, since they are not very
illuminating. We finally note that the field equations of the scalar
fields $\Theta$ give rise to a duality relation between the 2-form
potentials $B$ and the embedding scalars $\Theta$
\cite{Bergshoeff:2008qd,deWit:2008ta}.

\section{Discussion}

In this note we have presented a derivation of the Bagger-Lambert
theory of multiple M2-branes by an application of the embedding
tensor method to ${\cal N}=8$ supersymmetric gauge theories in three
dimensions. The linear constraint imposed by global supersymmetry
restricts the embedding tensor to an anti-symmetric 4-index tensor,
giving rise to the Bagger-Lambert theory. This is in contrast to the
case of ${\cal N}=8$ supergravity, where the linear constraint also
allows for a symmetric traceless tensor and a singlet
\cite{deWit:2003ja}. These representations lead to extra gaugings in
the supergravity case. For instance, for  $f_{abcd}=0$ a consistent
gauging is obtained by the embedding of the compact gauge group
$SO(p)\times SO(N-p)$ into $SO(N)$ with opposite coupling constant
for the two different groups \cite{deWit:2003ja}. We hope that the
relation with gauged supergravities can be helpful in finding more
solutions to the quadratic constraints.

In a second stage we have replaced the embedding tensor $\Theta$ by
scalar fields $\Theta (x)$. This has several advantages. First of
all the Chern-Simons terms can now be made manifestly invariant
under parity transformations by taking the scalars $\Theta$ to be
odd under parity. Secondly, the theory contains less free
parameters: the constants $\Theta$ have become integration constants
that occur only after solving the equations of motion. Thirdly, the
scalars $\Theta$ allow the possibility of domain walls on the
M2-brane worldvolume where, upon crossing the domain wall, the
constants $\Theta$ change value \cite{Bergshoeff:1998ys}.

It is of interest to search for generalizations of the
Bagger-Lambert model. For a recent discussion, see
\cite{Morozov:2008cb}. Another promising approach is to consider
supersymmetric gauge theories without a Lagrangian
\cite{Gran:2008vi}. In fact, gauged supergravities without a
Lagrangian have already been considered in the literature, see,
e.g., \cite{Bergshoeff:2003ri}. We expect that the application of
the embedding tensor technique in these cases will lead to more
general gaugings.

Quite a few  papers have appeared recently addressing different
issues concerning the world-volume theory of multiple M2-branes. In
particular, the relation with multiple D2-branes has been clarified
\cite{Mukhi:2008ux} (see also \cite{Gran:2008vi}), the $OSp(8|4)$
superconformal symmetry of the model has been verified
\cite{Bandres:2008vf}, the boundary theory of open membranes has
been considered \cite{Berman:2008be} and it has been shown that the
$SO(4)$ gauge theory solution corresponds to two  M2-branes moving
on a non-trivial manifold \cite{Lambert:2008et,Distler:2008mk}. We
hope that  this note will help in further clarifying the relation
between (and possible extensions of) ${\cal N}=8$ superconformal
theories and multiple M2-branes.

\subsection*{Acknowledgments}
E.B. would like to thank Diederik Roest for very insightful
discussions which led to the contents of section 3. He also
acknowledges stimulating discussions with the members of the journal
club of the physics department of Barcelona University.
 This work was partially supported by the European
Commission FP6 program MRTN-CT-2004-005104EU and by the INTAS
Project 1000008-7928.

\end{document}